\documentclass[12pt]{article}
\usepackage{amsmath}
\usepackage{amssymb}
\usepackage{epsfig}
\usepackage{euscript}
\usepackage{fancybox}
\usepackage{color}
\def\Color#1{\color[named]{#1}}

\def\mathswitchr#1{\relax\ifmmode{\mathrm{#1}}\else$\mathrm{#1}$\fi}

\newcommand{\umf}{{\Color{PineGreen}\mathfrak{u}}}
\newcommand {\pslash}{\hbox{$\not\hbox{\kern-2.3pt $p$}$}}

\usepackage{cite}

\usepackage{epic}

\def\alf1{ {\alpha\over\pi} }

\begin{document}
\begin{titlepage}
\begin{flushright}
{\bf BU-HEPP-05-06 }\\
{\bf July, 2005}\\
\end{flushright}
 
\begin{center}
{\Large IR-Improved DGLAP Theory$^{\dagger}$
}
\end{center}

\vspace{2mm}
\begin{center}
{\bf   B.F.L. Ward}\\
\vspace{2mm}
{\em Department of Physics,\\
 Baylor University, Waco, Texas, 76798-7316, USA}\\
\end{center}

\vspace{5mm}
\begin{center}
{\bf   Abstract}
\end{center}
We show that it is possible to improve the infrared aspects of
the standard treatment of the DGLAP evolution theory to take into account
a large class of higher order corrections that significantly improve the
precision of the theory for any given level of fixed-order
calculation of its respective kernels. We illustrate the size of the effects
we resum using the moments of the parton distributions.\par 
\vspace{10mm}
\vspace{10mm}
\renewcommand{\baselinestretch}{0.1}
\footnoterule
\noindent
{\footnotesize
\begin{itemize}
\item[${\dagger}$]
Work partly supported by US DOE grant DE-FG02-05ER41399 and
by NATO grant PST.CLG.980342.
\end{itemize}
}

\end{titlepage}

\def\Kmax{K_{\rm max}}\def\ieps{{i\epsilon}}\def\rQCD{{\rm QCD}}
\renewcommand{\theequation}{\arabic{equation}}
\font\fortssbx=cmssbx10 scaled \magstep2
\renewcommand\thepage{}
\parskip.1truein\parindent=20pt\pagenumbering{arabic}\par
In the preparation of the physics for the precision 
QCD$\times$EW(electroweak)~\cite{sm1,qcd}
LHC physics studies, all aspects of the calculation of the cross sections
and distributions for the would-be physical observables must be re-examined
if precision tags such as that envisioned for the luminosity theoretical
precision are to be realized, i.e., 1\% cross section predictions for 
single heavy gauge boson production
in 14 TeV pp collisions when that heavy gauge boson decays
into a light lepton pair. The QCD DGLAP\cite{dglap}
evolution of the structure functions
from the typical reference scale of data input, $\mu_0\sim 1-2GeV$,
to the respective hard scale
is one step that warrants further study, as it is well-known to many.
Many authors~\cite{cteq,mrst,greya,other-pdfs} have provided excellent
realizations of this evolution in the recent literature.
Here, we will re-examine the infrared aspects of the basic DGLAP theory itself
to try to improve the treatment to a level consistent with
the new era of precision QCD$\times$EW physics needed for the LHC physics
objectives.\par
\par

Specifically, the motivation for the improvement which we develop
can be seen already in the basic results in Refs.~\cite{dglap}
for the kernels that determine the evolution of the structure
functions by the attendant DGLAP evolution of the corresponding
parton densities by the standard methodology. Consider the evolution of the
non-singlet(NS) parton density function $q^{NS}(x)$, where $x$
can be identified as Bjorken's variable as usual.
The basic starting point of our analysis is the infrared divergence
in the kernel that determines this evolution:
\begin{equation}
\frac{dq^{NS}(x,t)}{dt}=\frac{\alpha_s(t)}{2\pi}\int_{x}^{1}\frac{dy}{y}q^{NS}(y,t)P_{qq}(x/y)
\label{dglap1}
\end{equation}
where the well-known result for the kernel $P_{qq}(z)$ is,
for $z<1$,
\begin{equation}
P_{qq}(z)= C_F\frac{1+z^2}{1-z}
\label{dglap2}
\end{equation}
when we set $t=\ln \mu^2/\mu_0^2$ for some reference scale $\mu_0$
with which we study evolution to the scale of interest $\mu$.
\footnote{We will generally follow Ref.~\cite{field} and set
$\mu_0=\Lambda_{QCD}$ without loss of content since $dt=dt'$
when $t=\ln\mu^2/\Lambda_{QCD}^2,~ t'=\ln\mu^2/\mu_0^2$ for fixed
values of $\Lambda_{QCD},\mu_0$.}
Here, $C_F=(N_c^2-1)/(2N_c)$ is the quark color 
representation's quadratic Casimir invariant
where $N_c$ is the number of colors and so that it is just 3. 
This kernel has an unintegrable IR singularity at $z=1$, which is the point of
zero energy gluon emission and this is as it should be.
The standard treatment of this very physical effect is to
regularize it by the replacement
\begin{equation}
\frac{1}{(1-z)}\rightarrow \frac{1}{(1-z)_+}
\label{dglap3}
\end{equation} 
with the distribution $\frac{1}{(1-z)_+}$ defined so that
for any suitable test function $f(z)$ we have
\begin{equation}
\int_{0}^{1}dz\frac{f(z)}{(1-z)_+}=\int_{0}^{1}dz\frac{f(z)-f(1)}{(1-z)}.
\label{dglap4}
\end{equation}
A possible representation of $1/(1-z)_+$ is seen to be
\begin{equation}
\frac{1}{(1-z)_+}=\frac{1}{(1-z)}\theta(1-\epsilon-z)+\ln\epsilon\,\delta(1-z)
\label{dglap5}
\end{equation}
with the understanding that $\epsilon\downarrow 0$.
We use the notation $\theta(x)$ for the step function from $0$ for $x<0$
to $1$ for $x\ge 0$ and $\delta(x)$ is Dirac's delta function.
The final result for $P_{qq}(z)$ is then obtained by imposing the
physical requirement~\cite{dglap} that
\begin{equation}
\int_{0}^{1}dzP_{qq}(z)=0,
\label{dglap6}
\end{equation}  
which is satisfied by adding the effects of virtual corrections at
$z=1$ so that finally
\begin{equation}
P_{qq}(z)= C_F\left( \frac{1+z^2}{(1-z)_+}+\frac{3}{2}\delta(1-z)\right). 
\label{dglap7}
\end{equation}  
\par

The smooth behavior in the original real emission 
result from the Feynman rules,
with a divergent $1/(1-z)$ behavior as $z\rightarrow 1$, has been replaced
with a mathematical artifact: the regime $1-\epsilon< z<1$ now has
no probability at all 
and at $z=1$ we have a large negative integrable contribution
so that we end-up finally with a finite (zero) value for the 
total integral of $P_{qq}(z)$. This mathematical artifact is what we wish
to improve here; for, in the precision studies of 
Z physics~\cite{yellowbook,jsw,jw} at LEP1,
it has been found that such mathematical artifacts can indeed impair
the precision tag which one can achieve with a given fixed order
of perturbation theory. An analogous case is now well-known in the theory of
QCD higher order
corrections, where the FNAL data on $p_T$ spectra clearly show the 
need for improvement of fixed-order results by resumming large logs associated with soft gluons~\cite{berge,cdf1}. For reference, note that at the LHC, 2 TeV
partons are realistic so that $z\cong 0.001$ means $\sim 2-3$ GeV soft gluons, which are clearly above the LHC detector thresholds, 
in complete analogy with the situation
at LEP where $z\cong 0.001$ meant $\sim 100$ MeV photons which 
were also above the LEP detector thresholds -- just as resummation 
was necessary to describe this view of the LEP data, so too we may 
argue it will be necessary to describe the LHC data on the corresponding view.
And, more importantly, why should we have
to set $P_{qq}(z)$ to $0$ for $1-\epsilon< z < 1$ when it actually has
its largest values in this very regime?
\par

By mathematical artifact we do not mean that there is an error 
in the computations that lead to it; indeed, it is well-known
that this +-function behavior is exactly what one gets at
${\cal O}(\alpha_s)$ for the bremsstrahlung process. The artifact
is that the behavior of the differential spectrum of 
the process for $z\rightarrow 1$ in ${\cal O}(\alpha_s)$
is unintegrable and has to be cut-off 
and thus this spectrum is only poorly represented by the ${\cal O}(\alpha_s)$ calculation; for, the resummation of the large soft higher order effects as 
we present below changes the $z\rightarrow 1$ behavior non-trivially,
as from our resummation we will find that the $\frac{1}{1-z}$-behavior
is modified to $(1-z)^{\gamma-1}, \gamma>0$.
This is a testable effect, as we have seen in its QED analogs in Z physics at
LEP1~\cite{yellowbook,jsw,jw}: 
if the experimentalist measures the cross section
for bremsstrahlung for gluons(photons) down to energy fraction $\epsilon_0,~\epsilon_0>0$, in our new resummed theory presented below, 
the result will approach
a finite value from below as $-\epsilon_0^\gamma$ whereas the ${\cal O}(\alpha_s)$ +-function prediction would increase without limit as $-\ln\epsilon_0$.
The exponentiated result has been verified by the data at LEP1.\par

The important point is that the traditional resummations in N-moment
space for the DGLAP kernels address only the short-distance 
contributions to their higher order corrections. The deep question we
deal with in this paper concerns, then, how much of the complete
soft limit of the DGLAP kernels is contained in the anomalous dimensions
of the leading twist operators in Wilson's expansion, an expansion
which resides on the very tip of the light-cone? Are all of the effects of
the very soft gluon emission, involving, as they most certainly do, arbitrarily
long wavelength quanta, representable by the physics
at the tip of the light-cone? The Heisenberg uncertainty principle
surely tells us that answer can not be affirmative. In this paper, we
calculate these long-wavelength gluon effects on the DGLAP kernels
that are not included (see the discussion below)
in the standard treatment of Wilson's expansion. We therefore do not
contradict the results of the large N-moment space resummations 
such as that presented in Ref.~\cite{alball} nor do we
contradict the renormalon chain-type resummation as done in
Ref.~\cite{mikhlv}.\par

We employ the exact re-arrangement of the Feynman series for QCD 
as it has been shown in Ref.~\cite{qcdexp,delaney}. For completeness,
as this QCD exponentiation theory is not generally familiar, we
reproduce its essential aspects in our Appendix. The idea is to sum up the
leading IR terms in the corrections to $P_{qq}$ with the goal that
they will render integrable the IR singularity that we have in its lowest order
form. This will remove the need for mathematical artifacts
and exhibit more accurately the true predictions of the 
full QCD theory in terms of fully
physical results.\par

\begin{figure}
\begin{center}
\setlength{\unitlength}{1mm}
\begin{picture}(160,80)
\put(-2.4, -10){\makebox(0,0)[lb]{
\epsfig{file=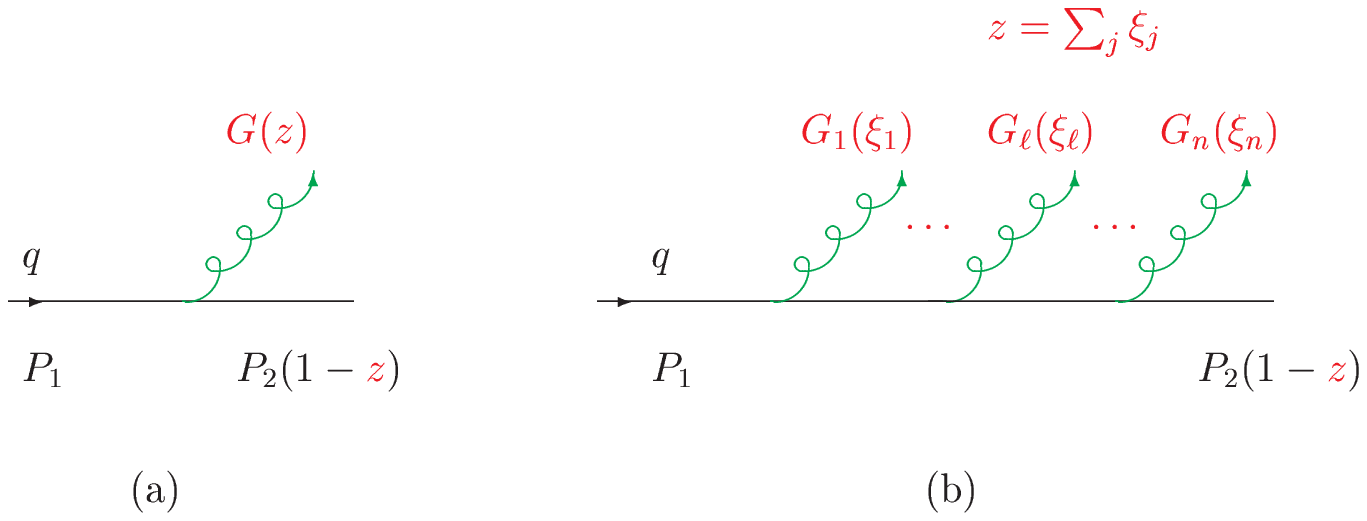,width=140mm}
}}
\end{picture}
\end{center}
\label{fig1-a}
\caption{In (a), we show the usual process $q\rightarrow q(1-z)+G(z)$;
in (b), we show its multiple gluon improvement $q\rightarrow q(1-z)+G_1(\xi_1)+\cdots+G_n(\xi_n),~~z=\sum_j\xi_j$.} 
\end{figure}

We apply the QCD exponentiation master formula in eq.(\ref{subp15})
in our Appendix
(see also Ref.~\cite{qcdexp}), following the
analogous discussion then for QED in Refs.~\cite{jsw,jw},
to the gluon emission transition that
corresponds to $P_{qq}(z)$, i.e., to the squared amplitude for
$q\rightarrow q(z)+G(1-z)$ so that in the Appendix one replaces
everywhere the squared amplitudes for the $\bar{Q}'Q\rightarrow \bar{Q}'''Q''$
processes with those for the former one plus its $nG$ analoga 
with the attendant changes in the
phase space and kinematics dictated by the standard methods; this implies
that in eq.(53) of the first paper in Ref.~\cite{dglap} we
have from eq.(\ref{subp15}) the replacement ( see Fig.~\ref{fig1-a} )
\noindent
\begin{equation}
\begin{split}
P_{BA}&=P_{BA}^0\equiv\frac{1}{2}z(1-z)\overline{\underset{spins}{\sum}}~\frac{|V_{A\rightarrow B+C}|^2}{p_\perp^2}\\
&\Rightarrow\\
P_{BA}&=\frac{1}{2}z(1-z)\overline{\underset{spins}{\sum}}~\frac{|V_{A\rightarrow B+C}|^2}{p_\perp^2}z^{\gamma_q}F_{YFS}(\gamma_q)e^{\frac{1}{2}\delta_q} 
\end{split}
\label{expn1-dglp}
\end{equation}
where $A=q$, $B=G$, $C=q$ and $V_{A\rightarrow B+C}$ is the lowest order
amplitude
for $q\rightarrow G(z)+q(1-z)$, so that 
we get the un-normalized exponentiated result
\begin{equation}
P_{qq}(z)= C_F F_{YFS}(\gamma_q)e^{\frac{1}{2}\delta_q}\frac{1+z^2}{1-z}(1-z)^{\gamma_q}
\label{dglap8}
\end{equation} 
where~\cite{qcdexp,delaney,jsw,jw} 
\begin{align}
\gamma_q &= C_F\frac{\alpha_s}{\pi}t=\frac{4C_F}{\beta_0}\\
\delta_q&=\frac{\gamma_q}{2}+\frac{\alpha_sC_F}{\pi}(\frac{\pi^2}{3}-\frac{1}{2})
\label{dglap9}
\end{align}
and 
\begin{equation}
F_{YFS}(\gamma_q)=\frac{e^{-C_E\gamma_q}}{\Gamma(1+\gamma_q)}.
\label{dglap10}
\end{equation}
Here, \[\beta_0=11-\frac{2}{3}n_f\], where $n_f$ is the number of
active quark flavors, \[C_E=.5772\dots\] is Euler's constant
and $\Gamma(w)$ is Euler's gamma function. The function
$F_{YFS}(z)$ was already introduced by Yennie, Frautschi
and Suura~\cite{yfs} in their analysis of the IR behavior
of QED. We see immediately that
the exponentiation has removed the unintegrable IR divergence at $z=1$.
For reference, we note that we have in (\ref{dglap8}) resummed
the terms\footnote{Following the standard LEP Yellow Book~\cite{yellowbook} convention, we do not include the order of the first nonzero term in
counting the order of its higher order corrections.} ${\cal O}(\ln^k(1-z)t^{\ell}\alpha_s^n),~~n\ge\ell\ge k$, which originate
in the IR regime and which exponentiate. The important point is that
we have not dropped outright the terms that do not exponentiate
but have organized them into the residuals $\tilde{\bar\beta}_m$
in the analog of eq.(\ref{subp15}).
   The application of eq.(\ref{subp15}) to obtain eq.(\ref{dglap8})
proceeds as follows. First, the exponent in the exponential factor in front of the expression on the RHS of eq.(\ref{subp15}) is readily seen to be from 
eq.(\ref{subp12}), using the well-known results for the respective
real and virtual infrared functions from Refs.~\cite{qcdexp,delaney},
\begin{equation}
\begin{split}
{SUM}_{IR}(QCD)&=2\alpha_s ReB_{QCD}+2\alpha_s\tilde B_{QCD}(\Kmax)\\
&=\frac{1}{2}\left(2 C_F\frac{\alpha_s}{\pi}t\ln{\frac{K_{max}}{E}}+ C_F\frac{\alpha_s}{2\pi}t+\frac{\alpha_sC_F}{\pi}(\frac{\pi^2}{3}-\frac{1}{2})\right)
\end{split}
\end{equation} 
where on the RHS of the last result we have already applied the DGLAP
synthesization procedure in the third paper in Ref.~\cite{qcdexp} 
to remove the collinear
singularities, $\ln \Lambda^2_{QCD}/m_q^2 -1 $, in accordance with the standard
QCD factorization theorems~\cite{qcdfactorzn}. This means that, identifying
the LHS of eq.(\ref{subp15}) as the sum over final states and average 
over initial states of the respective process divided by the incident
flux and replacing that incident flux by the respective initial state
density according to the standard methods for the
process $q\rightarrow q(1-z)+G(z)$, occurring in the context
of a hard scattering at scale $Q$ as it is for eq.(53) in the first paper
in Ref.~\cite{dglap}, the soft gluon effects for energy fraction
$<z\equiv K_{max}/E$
give the result, from eq.(\ref{subp15}), that, working through to
the $\tilde{\bar\beta_1}$-level and using $q_2$ to represent the momentum
conservation via the other degrees of freedom for the attendant hard process, 
\begin{equation}
\begin{split}
\int\frac{\alpha_s(t)}{2\pi}P_{BA}dtdz&=e^{\rm SUM_{IR}(QCD)(z)}\int\{\tilde{\bar\beta}_0\int{d^4y\over(2\pi)^4}e^{\{iy\cdot(p_1-p_2)+\int^{k<K_{max}}{d^3k\over k}\tilde S_\rQCD(k)
\left[e^{-iy\cdot k}-1\right]\}}\\
&+ \int{d^3
k_1\over k_1}\tilde{\bar\beta}_1(k_1)\int{d^4y\over(2\pi)^4}e^{\{iy\cdot(p_1-p_2-k_1)+\int^{k<K_{max}}{d^3k\over k}\tilde S_\rQCD(k)
\left[e^{-iy\cdot k}-1\right]\}}\\
&+\cdots \}{d^3p_2\over p_2^{\,0}}{d^3q_2\over q_2^{\,0}} \\
&= e^{\rm SUM_{IR}(QCD)(z)}\int\{\tilde{\bar\beta}_0\int_{-\infty}^{\infty}{dy\over(2\pi)}e^{\{iy\cdot(E_1-E_2)+\int^{k<K_{max}}{d^3k\over k}\tilde S_\rQCD(k)
\left[e^{-iyk}-1\right]\}}\\
&+ \int{d^3
k_1\over k_1}\tilde{\bar\beta}_1(k_1)\int_{-\infty}^{\infty}{dy\over(2\pi)}e^{\{iy\cdot(E_1-E_2-k_1^0)+\int^{k<K_{max}}{d^3k\over k}\tilde S_\rQCD(k)
\left[e^{-iy\cdot k}-1\right]\}}\\
&+\cdots \}{d^3p_2\over p_2^{\,0}q_2^{\,0}} 
\end{split}
\label{expker1}
\end{equation}
where we set $E_i=p_i^0,~ i=1,2$ and the real infrared function $\tilde S_\rQCD(k)$ is well-known as well:
\begin{equation}
\tilde S_\rQCD(k)= -\frac{\alpha_s C_F}{8\pi^2}\left(\frac{p_1}{kp_1}-\frac{p_2}{kp_2}\right)^2|_{\text{DGLAP synthesized}}
\label{realir}
\end{equation}
and we indicate as above that the DGLAP synthesization
procedure in Refs.~\cite{qcdexp} is to be applied to its evaluation to
remove its collinear singularities; we are using the 
kinematics of the first paper in Ref.~\cite{dglap} in their
computation of $P_{BA}(z)$ in their eq.(53), so that the 
relevant value of $k_\perp^2$ is indeed $Q^2$. It means that  
the computation can also be seen to correspond to computing the IR 
function for the standard t-channel kinematics and taking $\frac{1}{2}$ 
of the result to match the single line emission in $P_{Gq}$. 
The two important integrals needed in (\ref{expker1}) were already
studied in Ref.~\cite{yfs}:
\begin{equation}
\begin{split}
I_{YFS}(zE,0)&=\int_{-\infty}^{\infty}\frac{dy}{2\pi}e^{[iy(zE)+\int^{k<zE}\frac{d^3k}{k}\tilde{S}_{QCD}(k)(e^{-iyk}-1)]}\\
&= F_{YFS}(\gamma_q)\frac{\gamma_q}{zE}\\
I_{YFS}(zE,k_1)&=\int_{-\infty}^{\infty}\frac{dy}{2\pi}e^{[iy(zE-k_1)+\int^{k<zE}\frac{d^3k}{k}\tilde{S}_{QCD}(k)(e^{-iyk}-1)]}\\
&=(\frac{zE}{zE-k_1})^{1-\gamma_q}I_{YFS}(zE,0)
\end{split}
\label{yfsintgls} 
\end{equation}
\par
When we introduce the results in (\ref{yfsintgls}) into (\ref{expker1})
we can identify the factor
\begin{equation}
\int\left(\tilde{\bar\beta}_0\frac{\gamma_q}{zE}+\int dk_1k_1d\Omega_1\tilde{\bar\beta}_1(k_1)(\frac{zE}{zE-k_1})^{1-\gamma_q}\frac{\gamma_q}{zE}\right)\frac{d^3p_2}{E_2q_2^{\,0}}=\int dt \frac{\alpha_s(t)}{2\pi}P_{BA}^0dz+{\cal O}(\alpha_s^2).
\label{1storder}
\end{equation}
where $P_{BA}^0$ is the unexponentiated result in the first line of
(\ref{expn1-dglp}). This leads us finally to the exponentiated 
result in the second line
of (\ref{expn1-dglp}) by elementary differentiation:
\begin{equation}
P_{BA}=P_{BA}^0z^{\gamma_q}F_{YFS}(\gamma_q)e^{\frac{1}{2}\delta_q}
\end{equation}
\par
Here, we also may note
how one can see that the terms we exponentiate are not included
in the standard treatment of Wilson's expansion: From the 
standard method~\cite{fyodor}, the N-th moment of the invariants $T_{i,\ell},~i=L,2,3,~\ell=q,G,$
of the forward Compton amplitude in DIS is projected by
\begin{equation}
{\cal P}_N\equiv \left[\frac{q^{\{\mu_1}\cdots q^{\mu_N\}}}{N!}\frac{\partial^N}{\partial{p^{\mu_1}}\cdots \partial{p^{\mu_N}}}\right]|_{p=0}
\end{equation}
where $x_{Bj}=Q^2/(2qp)$ in the standard DIS notation; this projects
the coefficient of $1/(2x_{Bj})^N$. For the dominant terms which we
resum here, the characteristic behavior would correspond formally
to $\gamma_q$-dependent anomalous dimensions associated with
the respective coefficient whereas by definition Wilson's
expansion does not contain such. In more phenomenologically familiar
language, it is well-known that the parton model used in this paper
to calculate the large distance effects that improve the kernels contains
such effects whereas Wilson's expansion does not: for example, the 
parton model can be used for Drell-Yan processes, Wilson's expansion
can not. Simalarly, any Wilson-expansion guided procedure
used to infer the kernels via inverse Mellin transformation,
by calculating the coefficient of $(1/z)^n$ in Wilson's expansion,
will necessarily omit the dominant IR terms which we resum.
Here, we stress that we refer to the properties of the 
expansion of the invariant functions $T_i$, 
{\em not to the expansion of the kernels themselves}, 
as the latter are related to the respective 
anomalous dimension matrix elements by inverse Mellin transformations.
\par
 
The normalization condition in eq.(\ref{dglap6}) then gives us the final
expression
\begin{equation}
P_{qq}(z)= C_F F_{YFS}(\gamma_q)e^{\frac{1}{2}\delta_q}\left[\frac{1+z^2}{1-z}(1-z)^{\gamma_q} -f_q(\gamma_q)\delta(1-z)\right]
\label{dglap11}
\end{equation} 
where
\begin{equation}
f_q(\gamma_q)=\frac{2}{\gamma_q}-\frac{2}{\gamma_q+1}+\frac{1}{\gamma_q+2}.
\label{dglap12}
\end{equation} 
The latter result is then our IR-improved kernel for NS DGLAP
evolution in QCD. We note that the appearance of the integrable
function $(1-z)^{-1+\gamma_q}$ in the place of $\frac{1}{(1-z)_+}$
was already anticipated by Gribov and Lipatov in Refs.~\cite{dglap}.
Here, we have calculated the value of $\gamma_q$ in a systematic
rearrangement of the QCD perturbation theory that allows 
one to work to any exact order in the theory without
dropping any part of the theory's perturbation series.\par

The standard DGLAP theory tells us that the kernel $P_{Gq}(z)$ is 
related to $P_{qq}(1-z)$ directly: for $z<1$, we have
\begin{equation}
P_{Gq}(z)=P_{qq}(1-z)= C_F F_{YFS}(\gamma_q)e^{\frac{1}{2}\delta_q}\frac{1+(1-z)^2}{z} z^{\gamma_q} .
\label{dglap13}
\end{equation}
This then brings us to our first non-trivial check of the new IR-improved theory; for, the conservation of momentum tells us that
\begin{equation}
\int_{0}^{1}dz z \left(P_{Gq}(z)+P_{qq}(z)\right) = 0.
\label{dglap14}
\end{equation}
Using the new results in eqs.(\ref{dglap11},\ref{dglap13}), we have
to check that the following integral vanishes:
\begin{equation}
I = \int_{0}^{1}dz z \left(\frac{1+(1-z)^2}{z} z^{\gamma_q}+\frac{1+z^2}{1-z}(1-z)^{\gamma_q} -f_q(\gamma_q) \delta(1-z)\right).
\label{dglap15}
\end{equation}
To see that it does, note that
\begin{equation}
\frac{z}{1-z}=\frac{z-1+1}{1-z}=-1+\frac{1}{1-z}.
\end{equation}
Introducing this result into eq.(\ref{dglap15}) we get
\begin{equation}
I = \int_{0}^{1}dz\{ (1+(1-z)^2)z^{\gamma_q}-(1+z^2)(1-z)^{\gamma_q}+\frac{1+z^2}{1-z}(1-z)^{\gamma_q} -f_q(\gamma_q) \delta(1-z)\}.
\label{dglap16}
\end{equation}
The integrals over the first two terms on the right-hand side (RHS)
of (\ref{dglap16}) exactly cancel as one sees by using the change
of variable $z\rightarrow 1-z$ in one of them and the integral
over the last two terms on the RHS of (\ref{dglap16}) vanishes from the
normalization in eq.(\ref{dglap6}). Thus we conclude that
\begin{equation}
I=0.
\end{equation}
The quark momentum sum rule is indeed satisfied.
\par

Having improved the IR divergence properties of $P_{qq}(z)$ and
$P_{Gq}(z)$, we now turn to $P_{GG}(z)$ and $P_{qG}(z)$.
We first note that the standard formula for $P_{qG}(z)$,
\begin{equation}
P_{qG}(z)=\frac{1}{2}(z^2+(1-z)^2),
\label{dglap17a}
\end{equation}
is already well-behaved (integrable) in the IR regime. Thus, we do not
need to improve it here to make it integrable and we
note that the singular contributions in the other
kernels are expected to dominate the evolution effects in any case. We do not exclude improving it for the best precision~\cite{elswh} and we
return to this
point presently.\par

This brings us then to $P_{GG}(z)$. Its lowest order form is
\begin{equation}
P_{GG}(z)= 2C_{G}(\frac{1-z}{z}+\frac{z}{1-z}+z(1-z))
\label{dglap17}
\end{equation} 
which again exhibits unintegrable IR singularities at both $z=1$ and $z=0$.
(Here, $C_G$ is the gluon quadratic Casimir invariant, so that it is just 
$N_c=3$.) If we repeat the QCD exponentiation calculation carried-out above
by using the color representation for the gluon rather than that for the
quarks, i.e., if we apply the exponentiation analysis in the Appendix to
the squared amplitude for the process $G\rightarrow G(z)+G(1-z)$, we get the exponentiated un-normalized result{\small
\begin{equation}
P_{GG}(z)= 2C_G F_{YFS}(\gamma_G)e^{\frac{1}{2}\delta_G}\left(\frac{1-z}{z}z^{\gamma_G}+\frac{z}{1-z}(1-z)^{\gamma_G}+\frac{1}{2}(z^{1+\gamma_G}(1-z)+z(1-z)^{1+\gamma_G})\right)
\label{dglap18}
\end{equation}}
wherein we obtain the $\gamma_G$ and $\delta_G$ from the expressions for
$\gamma_q$ and $\delta_q$ by the substitution $C_F\rightarrow C_G$:
\begin{align}
\gamma_G &= C_G\frac{\alpha_s}{\pi}t=\frac{4C_G}{\beta_0}\\
\delta_G&=\frac{\gamma_G}{2}+\frac{\alpha_sC_G}{\pi}(\frac{\pi^2}{3}-\frac{1}{2}).
\label{dglap19}
\end{align}
We see again that exponentiation has again made the singularities at 
$z=1$ and $z=0$ integrable.\par

To normalize $P_{GG}$, we take into account the virtual corrections such that
the gluon momentum sum rule  
\begin{equation}
\int_{0}^{1}dz z \left(2n_f P_{qG}(z)+P_{GG}(z)\right) = 0
\label{dglap20}
\end{equation}
is satisfied.
This gives us finally the IR-improved result
\begin{equation}
\begin{split}
P_{GG}(z)&= 2C_G F_{YFS}(\gamma_G)e^{\frac{1}{2}\delta_G}\{ \frac{1-z}{z}z^{\gamma_G}+\frac{z}{1-z}(1-z)^{\gamma_G}\\
&\qquad +\frac{1}{2}(z^{1+\gamma_G}(1-z)+z(1-z)^{1+\gamma_G}) - f_G(\gamma_G) \delta(1-z)\}
\end{split}
\label{dglap21}
\end{equation}
where for $f_G(\gamma_G)$ we get
\begin{equation}
\begin{split}
f_G(\gamma_G)&=\frac{n_f}{6C_GF_{YFS}(\gamma_G)}{e^{-\frac{1}{2}\delta_G}}+
\frac{2}{\gamma_G(1+\gamma_G)(2+\gamma_G)}+\frac{1}{(1+\gamma_G)(2+\gamma_G)}\\
&\qquad +\frac{1}{2(3+\gamma_G)(4+\gamma_G)}+\frac{1}{(2+\gamma_G)(3+\gamma_G)(4+\gamma_G)}.
\end{split}
\end{equation}
It is these improved results in eqs.(\ref{dglap11},\ref{dglap13},\ref{dglap21})
for $P_{qq}(z)$, $P_{Gq}(z)$ and $P_{GG}(z)$ that we 
use together with the standard result in (\ref{dglap17}) for
$P_{qG}(z)$ as the IR-improved DGLAP theory.\par

For clarity we summarize at this point 
the new IR-improved kernel set as follows:
\begin{align}
P_{qq}^{exp}(z)&= C_F F_{YFS}(\gamma_q)e^{\frac{1}{2}\delta_q}\left[\frac{1+z^2}{1-z}(1-z)^{\gamma_q} -f_q(\gamma_q)\delta(1-z)\right],\\
P_{Gq}^{exp}(z)&= C_F F_{YFS}(\gamma_q)e^{\frac{1}{2}\delta_q}\frac{1+(1-z)^2}{z} z^{\gamma_q},\\
P_{GG}^{exp}(z)&= 2C_G F_{YFS}(\gamma_G)e^{\frac{1}{2}\delta_G}\{ \frac{1-z}{z}z^{\gamma_G}+\frac{z}{1-z}(1-z)^{\gamma_G}\nonumber\\
&\qquad +\frac{1}{2}(z^{1+\gamma_G}(1-z)+z(1-z)^{1+\gamma_G}) - f_G(\gamma_G) \delta(1-z)\},\\
P_{qG}(z)&=\frac{1}{2}(z^2+(1-z)^2),
\label{dglap22}
\end{align}
where we have introduced the superscript $exp$ to denote the exponentiated
results henceforth.\par

Returning now to the improvement of $P_{qG}(z)$, let us record it as well
for the sake of completeness and of providing better precision. 
Applying eq.(\ref{subp15}) to the
process $G\rightarrow q+\bar{q}$, we get the exponentiated result
\begin{equation}
P_{qG}^{exp}(z)= F_{YFS}(\gamma_G)e^{\frac{1}{2}\delta_G}\frac{1}{2}\{ z^2(1-z)^{\gamma_G}+(1-z)^2z^{\gamma_G}\}.
\label{dglap221}
\end{equation}
The gluon momentum sum rule then gives the new normalization constant for the
$P_{GG}^{exp}$ via the result
\begin{equation}
\begin{split}
\bar{f}_G(\gamma_G)&=\frac{n_f}{C_G}\frac{1}{(1+\gamma_G)(2+\gamma_G)(3+\gamma_G)}+
\frac{2}{\gamma_G(1+\gamma_G)(2+\gamma_G)}+\frac{1}{(1+\gamma_G)(2+\gamma_G)}\\
&\qquad +\frac{1}{2(3+\gamma_G)(4+\gamma_G)}+\frac{1}{(2+\gamma_G)(3+\gamma_G)(4+\gamma_G)}.
\end{split}
\label{dglap222}
\end{equation} 
The constant $\bar{f}_G$ should be substituted for $f_G$ in $P_{GG}^{exp}$
whenever the exponentiated result in (\ref{dglap221}) is used.
These results (\ref{dglap22}),~(\ref{dglap221}), and ~(\ref{dglap222})
are our new improved DGLAP kernel set, with the option
exponentiating $P_{qG}$ as well.
Let us now look into their effects on the moments of the structure
functions by discussing the corresponding effects on the moments
of the parton distributions.\par

We know that moments of the kernels determine the exponents in the logarithmic
variation~\cite{dglap,qcd1} of the moments of the quark distributions
and, thereby, of the moments of the structure functions themselves.
To wit, in the non-singlet case, we have 
\begin{equation}
\frac{dM^{NS}_n(t)}{dt}=\frac{\alpha_s(t)}{2\pi}A^{NS}_nM^{NS}_n(t)
\label{dglap23}
\end{equation}
where
\begin{equation}
M^{NS}_n(t)=\int^1_0dz z^{n-1}q^{NS}(z,t)
\label{dglap24}
\end{equation}
and the quantity $A^{NS}_n$ is given by
\begin{align}
A^{NS}_n&=\int^1_0dz z^{n-1}P_{qq}^{exp}(z),\nonumber\\
        &= C_F F_{YFS}(\gamma_q)e^{\frac{1}{2}\delta_q}[B(n,\gamma_q)+B(n+2,\gamma_q)-f_q(\gamma_q)]
\label{dglap25}
\end{align}
where $B(x,y)$ is the beta function
given by \[B(x,y)=\Gamma(x)\Gamma(y)/\Gamma(x+y)\].
This should be compared to the un-IR-improved result~\cite{dglap,qcd1}:
\begin{equation}
A^{NS^o}_n\equiv C_F\left[-\frac{1}{2}+\frac{1}{n(n+1)}-2\sum_{j=2}^{n}\frac{1}{j}\right].
\label{dglap26}
\end{equation}
The asymptotic behavior for large $n$ is now very different, as
the IR-improved exponent approaches a constant, a multiple of $-f_q$,
as we would expect as $n\rightarrow \infty$ because
$\lim_{n\rightarrow \infty}z^{n-1} = 0$ for $0\le z<1$ 
whereas, as it is well-known,
the un-IR-improved result in (\ref{dglap26}) diverges as $-2C_F\ln n$
as $n\rightarrow \infty$. The two results are also different at finite
$n$: for $n=2$ we get, for example, for $\alpha_s\cong .118$~\cite{siggi},
\begin{equation}
A^{NS}_2 =
\begin{cases}
C_F(-1.33)&,~~\text{ un-IR-improved}\\
C_F(-0.966)&,~~\text{IR-improved}
\end{cases} 
\label{dglap27}
\end{equation}
so that the effects we have calculated are important for all $n$
in general. For completeness, we note that the solution to (\ref{dglap23})
is given by the standard methods as
\begin{equation}
\begin{split}
M^{NS}_n(t)&=M^{NS}_n(t_0)e^{\int_{t_0}^{t}dt'\frac{\alpha_s(t')}{2\pi}A^{NS}_n(t')}\\
&=M^{NS}_n(t_0)e^{\bar{a}_n[Ei(\frac{1}{2}\delta_1\alpha_s(t_0))-Ei(\frac{1}{2}\delta_1\alpha_s(t))]} \\
&\qquad \operatornamewithlimits{\Longrightarrow}_{\small t,t_0~\text{large with}~t>>t_0}M^{NS}_n(t_0)\left(\frac{\alpha_s(t_0)}{\alpha_s(t)}\right)^{\bar{a'}_n}
\end{split}
\label{dglap27a} 
\end{equation}
where $Ei(x)=\int_{-\infty}^xdre^r/r$ is the exponential integral function,
\begin{equation}
\begin{split}
\bar{a}_n&=\frac{2C_F}{\beta_0}F_{YFS}(\gamma_q)e^{\frac{\gamma_q}{4}}[B(n,\gamma_q)+B(n+2,\gamma_q)-f_q(\gamma_q)]\\
\bar{a'}_n&=\bar{a}_n\left(1+\frac{\delta_1}{2}\frac{(\alpha_s(t_0)-\alpha_s(t))}{\ln(\alpha_s(t_0)/\alpha_s(t))}\right)
\end{split}
\label{dglap27b}
\end{equation}
with
\[\delta_1=\frac{C_F}{\pi}\left(\frac{\pi^2}{3}-\frac{1}{2}\right)\].
We can compare with the un-IR-improved result in which the last line
in eq.(\ref{dglap27a}) holds exactly with $\bar{a'}_n=2A^{NS^o}_n/\beta_0$.
Phenomenologically, for $n=2$, taking $Q_0=2$GeV and evolving to $Q=100$GeV,
if we set $\Lambda_{QCD}\cong .2GeV$ and use $n_f=5$ for definiteness of illustration, we see from eqs. (\ref{dglap27a},\ref{dglap27b}) 
that we get a shift of the 
respective evolved NS moment by $\sim~5\%$, 
which is of some interest in view of the
expected HERA precision~\cite{carli}.\par

We give now the remaining elements of the anomalous dimension
matrix in its 'best' IR-improved form for completeness:
\begin{align}
A^{Gq}_n&=\int_0^1dz z^{n-1} P^{exp}_{Gq}(z)= C_F F_{YFS}(\gamma_q)e^{\frac{1}{2}\delta_q}
\left[\frac{1}{n+\gamma_q-1}+B(3,n+\gamma_q-1)\right],\\
A^{GG}_n&=\int_0^1dz z^{n-1}P^{exp}_{GG}(z)= 2C_G F_{YFS}(\gamma_G)e^{\frac{1}{2}\delta_G}\{ B(n+1,\gamma_G)+B(n+\gamma_G-1,2)\nonumber\\
&\qquad +\frac{1}{2}(B(n+1,\gamma_G+2)+B(n+\gamma_G+1,2))-\bar{f}(\gamma_G)\},\\
2n_fA^{qG}_n&=2n_f\int^1_0dz z^{n-1}P^{exp}_{qG}(z)=2T(F)F_{YFS}(\gamma_G)e^{\frac{1}{2}\delta_G}\left(B(n+2,1+\gamma_G)+B(n+\gamma_G,3)\right),
\label{dglap28}
\end{align}
where $T(F)=\frac{1}{2}n_f$. We 
note that the un-exponentiated value of the 
last result in eq.(\ref{dglap28}) is
a well-known one~\cite{dglap,qcd1}, $2T(F)\frac{2+n+n^2}{n(n+1)(n+2)}$,
and it would be used whenever we do not choose to exponentiate $P_{qG}$.
We will investigate the further implications of these IR-improved
results for LHC physics elsewhere~\cite{elswh}.\par

In the discussion so far, we have used the lowest order DGLAP
kernel set to illustrate how important the resummation which we present
here can be. In the literature~\cite{mvermn,mvovermn}, there are now exact results up to ${\cal O}(\alpha_s^3)$ for the DGLAP kernels. The question naturally arises as to the relationship of our work to these fixed-order exact results.
We stress first that we are presenting an improvement of the fixed-order
results such that the singular pieces of the any exact fixed-order
result, i.e., the $\frac{1}{(1-z)_+}$ parts, are exponentiated so that they are
replaced with integrable functions proportional to $(1-z)^{\gamma-1}$ 
with $\gamma$ positive as we have illustrated above. Since the series
of logs which we resum to accomplish this has the structure
$\alpha_s^\ell t^\ell\ln^n(1-z),~\ell\ge n$
these terms are not already present
in the results in Refs.~\cite{mvermn,mvovermn}. As we use the formula
in eq.(\ref{subp15}), there will be no double counting if we
implement our IR-improvement of the exact fixed-order results in
Refs.~\cite{mvermn,mvovermn}. The detailed discussion of the
application of our theory to the results in Refs.~\cite{mvermn,mvovermn}
will appear elsewhere~\cite{elswh}. For reference, we note that the
higher order kernel corrections in Refs.~\cite{mvermn,mvovermn} are
perturbatively related to the leading order kernels, so one can expect that
the size
of the exponentiation effects illustrated above will only be perturbatively
modified by the higher order kernel corrections, leaving the same
qualitative behavior in general.
\par

In the interest of specificness, let us illustrate the IR-improvement
of $P_{qq}$ when calculated to three loops using the results
in Refs.~\cite{mvermn,mvovermn}. Considering the non-singlet case
for definiteness (a similar analysis holds for the singlet case)
we write in the notation of the latter references
\begin{equation}
P_{ns}^{+}=P_{qq}^v+P_{q\bar{q}}^v\equiv \sum_{n=0}^\infty(\frac{\alpha_s}{4\pi})^{n+1}P_{ns}^{(n)+}
\label{vermn1}
\end{equation}
where at order ${\cal O}(\alpha_s)$ we have 
\begin{equation}
P_{ns}^{(0)+}(z)=2C_F\{\frac{1+z^2}{(1-z)_+}+\frac{3}{2}\delta(1-z)\}
\label{vermn2}
\end{equation}
which shows that $P_{ns}^{(0)+}(z)$ agrees with the unexponentiated result in
(\ref{dglap7}) for $P_{qq}$ except for an overall factor of 2.
We use this latter identification to connect our work with that
in Refs.~\cite{mvermn,mvovermn} in the standard methodology.
In Refs.~\cite{mvermn,mvovermn}, exact results are given for
$P_{ns}^{(1)+}(z)$, and in Refs.~\cite{mvovermn} exact results are given
for $P_{ns}^{(2)+}(z)$.
When we apply the result in (\ref{subp15})
to the squared amplitudes for the processes $q\rightarrow q+X$, $\bar{q}\rightarrow q+X'$, we get the exponentiated result
\begin{equation}
\begin{split}
P_{ns}^{+,exp}(z)&=(\frac{\alpha_s}{4\pi}) 2P_{qq}^{exp}(z)+F_{YFS}(\gamma_q)e^{\frac{1}{2}\delta_q}\big{[}(\frac{\alpha_s}{4\pi})^2\{(1-z)^{\gamma_q}\bar{P}_{ns}^{(1)+}(z)+\bar{B}_2\delta(1-z)\}\\
&\qquad +(\frac{\alpha_s}{4\pi})^3\{(1-z)^{\gamma_q}\bar{P}_{ns}^{(2)+}(z)+\bar{B}_3\delta(1-z)\}\big{]}
\end{split}
\label{vermn3}
\end{equation} 
where $P_{qq}^{exp}(z)$ is given in (\ref{dglap22}) and the resummed residuals 
$\bar{P}_{ns}^{(i)+}$,~$i=1,2$ are related to the exact results for
$P_{ns}^{(i)+}$,~$i=1,2$, as follows:
\begin{equation}
\bar{P}_{ns}^{(i)+}(z)=P_{ns}^{(i)+}(z)-B_{1+i}\delta(1-z)+\Delta_{ns}^{(i)+}(z)
\label{vermn4}
\end{equation}
where 
\begin{equation}
\begin{split}
\Delta_{ns}^{(1)+}(z)&=-4C_F\pi\delta_1\{\frac{1+z^2}{1-z}-f_q\delta(1-z)\}\\
\Delta_{ns}^{(2)+}(z)&=-4C_F(\pi\delta_1)^2\{\frac{1+z^2}{1-z}-f_q\delta(1-z)\}\\
 &\qquad \qquad -2\pi\delta_1\bar{P}_{ns}^{(1)+}(z)
\end{split}
\label{vermn5}
\end{equation}
and
\begin{equation}
\begin{split}
\bar{B}_2&=B_2+4C_F\pi\delta_1f_q\\
\bar{B}_3&=B_3+4C_F(\pi\delta_1)^2f_q-2\pi\delta_1\bar{B}_2.
\end{split}
\label{vermn6}
\end{equation}
Here, the constants $B_i,~i=2,3$ are given by the results in
Refs.~\cite{mvermn,mvovermn} as
\begin{equation}
\begin{split}
B_2&=4C_GC_F(\frac{17}{24}+\frac{11}{3}\zeta_2-3\zeta_3)-4C_Fn_f(\frac{1}{12}+\frac{2}{3}\zeta_2)+4C_F^2(\frac{3}{8}-3\zeta_2+6\zeta_3)\\
B_3&=16C_GC_Fn_f(\frac{5}{4}-\frac{167}{54}\zeta_2+\frac{1}{20}\zeta_2^2+\frac{25}{18}\zeta_3)\\
&\qquad +16C_GC_F^2(\frac{151}{64}+\zeta_2\zeta_3-\frac{205}{24}\zeta_2-\frac{247}{60}\zeta_2^2+\frac{211}{12}\zeta_3+\frac{15}{2}\zeta_5)\\
&\qquad +16C_G^2C_F(-\frac{1657}{576}+\frac{281}{27}\zeta_2-\frac{1}{8}\zeta_2^2-\frac{97}{9}\zeta_3+\frac{5}{2}\zeta_5)\\
&\qquad +16C_Fn_F^2(-\frac{17}{144}+\frac{5}{27}\zeta_2-\frac{1}{9}\zeta_3)\\
&\qquad +16C_F^2n_F(-\frac{23}{16}+\frac{5}{12}\zeta_2+\frac{29}{30}\zeta_2^2-\frac{17}{6}\zeta_3)\\
&\qquad +16C_F^3(\frac{29}{32}-2\zeta_2\zeta_3+\frac{9}{8}\zeta_2+\frac{18}{5}\zeta_2^2+\frac{17}{4}\zeta_3-15\zeta_5),
\end{split}
\end{equation}
where $\zeta_n$ is the Riemann zeta function evaluated at argument $n$. 
The detailed phenomenological consequences of the fully exponentiated
2- and 3-loop DGLAP kernel set will appear elsewhere~\cite{elswh}.
\par
In summary, we have used exact re-arrangement of the QCD Feynman
series to isolate and resum the leading IR contributions
to the physical processes that generate the evolution kernels
in DGLAP theory. In this way, we have obviated the need to
employ artificial mathematical regularization of the attendant
IR singularities as the theory's higher order corrections
naturally tame these singularities. The resulting IR-improved 
anomalous dimension
matrix behaves more physically for large $n$ and receives significant
effects at finite $n$ from the exponentiation.\par

We in principle
can make contact with the moment-space resummation results
in Ref.~\cite{moment} but we stress that these results have necessarily been
obtained after making a Mellin transform of the mathematical
artifact which we address in this paper. Thus, the results
in Ref.~\cite{moment} do not in any way contradict the analysis 
in this paper.\par 

We note that the program of improvement of the hadron cross section
calculations for LHC physics advanced herein should be distinguished 
from the results in Refs.~\cite{sterman,cattrent}. Indeed, recalling the
standard hadron cross section formula
\begin{equation}
\sigma =\sum_{i,j}dx_1dx_2F_i(x_1)F_j(x_2)\hat\sigma(x_1x_2s)
\label{sigtot}
\end{equation}
where $\{F_\ell(x)\}$ are the respective parton densities and
$\hat\sigma(x_1x_2s)$ is the respective reduced hard parton cross section,
the resummation results in Refs.~\cite{sterman,cattrent} address,
by summing the large logs in Mellin transform space, 
the $x_1x_2\rightarrow 1$ limit of $\hat\sigma(x_1x_2s)$
whereas the results above address the improvement, by resummation
in x-space, of the calculation of the parton densities $\{F_i(x)\}$ for all
values of x. Thus, the program of improvement presented above is entirely 
complementary to that in Refs.~\cite{sterman,cattrent} and 
both programs of improvement are needed for precision LHC physics.\par

Finally, we address the issue of the relationship between the re-arrangement 
that we have made of the exact leading-logs in the QCD perturbation theory
and the usual treatment in the non-exponentiated DGLAP theory. If one
expands out the exponentiated kernels, using the distribution identity
\begin{equation}
(1-z)^{a-1}=\frac{1}{a}\delta(1-z)+\frac{1}{(1-z)_+}+\sum_{j=1}^{\infty}\frac{a^j}{j!}\left[\frac{\ln^j(1-z)}{1-z}\right]_+,
\label{dist1}
\end{equation} 
one can see that for example $P_{qq}$ and $P^{exp}_{qq}$ agree 
to leading order, so that the leading log series which they 
generate for the respective NS
structure functions also agree through leading order in $\frac{\alpha_s}{\pi}L$
where $L$ is the respective big log in momentum-space. At higher orders then,
we have a different result for the $\{F_i\}$, let us denote them 
by $\{{F'}_i\}$, and a different result for the reduced cross section, 
let us denote it by $\hat\sigma'$, such that we get the 
same perturbative QCD cross section,
\begin{equation}
\begin{split}
\sigma &=\sum_{i,j}\int dx_1dx_2F_i(x_1)F_j(x_2)\hat\sigma(x_1x_2s)\cr
       &=\sum_{i,j}\int dx_1dx_2{F'}_i(x_1){F'}_j(x_2)\hat\sigma'(x_1x_2s)
\end{split}
\end{equation}
order by order in perturbation theory. The exponentiated kernels are
used to factorize the mass singularities from the unfactorized reduced
cross section and this generates $\hat\sigma'$ instead of the usual
$\hat\sigma$ whose factorized form is generated using the usual DGLAP kernels.
We thus have the same leading log series for $\sigma$ as does the usual
calculation with un-exponentiated DGLAP kernels. We have an important
advantage: the lack of +-functions in the generation of the 
configuration space functions
$\{{F'}_i,\hat\sigma'\}$ means that these functions lend themselves 
more readily to Monte Carlo realization to arbitrarily soft radiative
effects, both for the generation of the parton shower associated
to the $\{{F'}_i\}$ and for the attendant remaining 
radiative effects in $\hat\sigma'$.
\par
Further consequences of our results
for LHC physics will be presented elsewhere\cite{elswh}. 

\section*{Acknowledgments}
We thank Prof. S. Jadach for useful discussions and Prof. W. Hollik
for the kind hospitality of the Max-Planck-Institut, Munich, wherein
a part of this work was completed.

\section*{Appendix}

In this Appendix, we present the new QCD exponentiation theory
which has been
developed in Refs.~\cite{qcdexp,delaney} as it is not generally
familiar. The goal is to make the current paper self-contained.\par

For definiteness,
we will use the process in Fig.~2, $\bar Q'(p_1) Q(q_1)\rightarrow \bar Q'''(p_2) Q''(q_2) +
G_1(k_1)\cdots G_n(k_n)$, as the proto-typical process, 
\begin{figure}
\begin{center}
\setlength{\unitlength}{1mm}
\begin{picture}(160,80)
\put(40, -8){\makebox(0,0)[lb]{
\epsfig{file=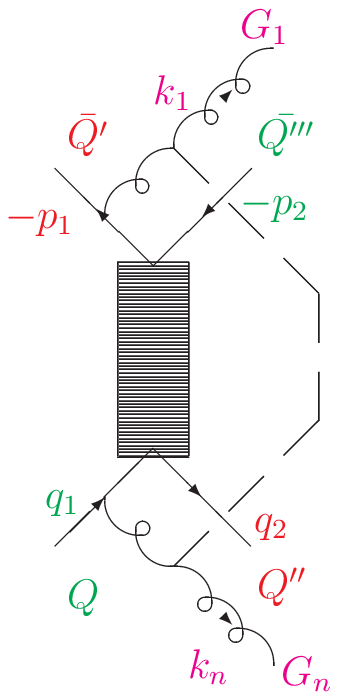,width=50mm}
}}
\end{picture}
\end{center}
\caption{\baselineskip=7mm     The process $\bar{Q'}
    Q      \rightarrow        \bar{Q'''}
                 +      Q''    + n(G)$.
The four--momenta are indicated in the standard manner: $q_1$ 
is the
four--momentum of the incoming $Q$, $q_2$ 
is the four--momentum of the outgoing $Q''$, etc.,
and $Q = u,d,s,c,b,G.$}
\label{figproto.1}
\end{figure}
\noindent
where we have written the
kinematics as it is illustrated in the figure. This process,
which dominates processes such as t\=t production at FNAL, contains all of the 
theoretical issues that we must face at the parton level to establish
, as an extension of the original 
ideas of Yennie, Frautschi and Suura (YFS)~\cite{yfs}, 
QCD soft exponentiation by MC methods -- applicability to other related
processes will be immediate. For reference, let us also 
note that, in what follows, we use the GPS conventions of 
Ref.~\cite{gps} for spinors $\{u,v,\umf\}$ and the attendant
photon and gluon polarization vectors that follow therefrom:
\begin{equation}
\label{gmapol}
  (\epsilon^\mu_\sigma(\beta))^*
     ={\bar{u}_\sigma(k) \gamma^\mu u_\sigma(\beta)
       \over \sqrt{2}\; \bar{u}_{-\sigma}(k) u_\sigma(\beta)},\quad
  (\epsilon^\mu_\sigma(\zeta))^*
     ={\bar{u}_\sigma(k) \gamma^\mu \umf_\sigma(\zeta)
       \over \sqrt{2}\; \bar{u}_{-\sigma}(k) \umf_\sigma(\zeta)},
\end{equation}
with $\beta^2=0$ and $\zeta$ defined in Ref.~\cite{gps}, 
so that all phase information is strictly known in our amplitudes.
This means that, although we shall use the older EEX realization of YFS
MC exponentiation as defined in Ref.~\cite{ceex:2001}, the realization
of our results via the the newer CEEX realization of YFS exponentiation
in Ref.~\cite{ceex:2001} is also possible and is in progress~\cite{elswh}.

Specifically, the authors in Refs.~\cite{delaney}
have analyzed how in the special case of Born level color exchange
one applies the YFS theory to QCD by extending
the respective YFS
IR singularity analysis to QCD to all orders in $\alpha_s$.
Here, unlike what was emphasized in Refs.~\cite{delaney},
we focus on the YFS theory as a general re-arrangement of
renormalized perturbation theory based on its IR
behavior, just as the renormalization group
is a general property of renormalized perturbation theory based on its
UV(ultra-violet) behavior.
We will thus keep our arguments
entirely general from the outset, so that it will be immediate
that our result applies to any renormalized perturbation theory
in which the cross section under study is finite.\par

Let the amplitude for the emission of $n$ real gluons in our proto-typical
subprocess, 
$Q^\alpha + {\bar {Q'}}^{\bar\alpha} \rightarrow {Q''}^\gamma
{\bar {Q'''}}^{\bar\gamma} + n(G)$, where $\alpha, {\bar\alpha}, \gamma$, and
${\bar\gamma}$ are color indices, be
represented by
\begin{equation}
{\cal M}^{(n)\alpha{\bar\alpha}}_{\gamma{\bar\gamma}} 
 = \sum_{\ell}M^{(n)\alpha{\bar\alpha}}_{\gamma{\bar\gamma}\ell},
\label{subp1}
\end{equation}
$M^{(n)}_{\ell}$ is the contribution to 
${\cal M}^{(n)}$
from Feynman diagrams with ${\ell}$ virtual loops.
Symmetrization yields
\begin{eqnarray}
   M^{(n)}_\ell = {1\over {\ell!}}\int\prod_{j=1}^{\ell}{d^4k_j\over{(2\pi)^4(
 k_j^2-\lambda^2+i\epsilon)}}\rho^{(n)}_\ell(k_1,\cdots,k_\ell),
\label{subp2}
\end{eqnarray}
where this last equation defines $\rho^{(n)}_\ell$ as a symmetric
function of its arguments arguments $k_1,...,k_{\ell}$.
$\lambda$ will be our infrared gluon regulator mass for IR singularities;
n-dimensional regularization of the 't Hooft-Veltman~\cite{tHvelt} type is also
possible as we shall see.

We now define the virtual IR emission factor $S_{QCD}(k)$ for a gluon
of 4-momentum $k$, for the $k\rightarrow 0$ regime of the
respective 4-dimensional loop integration as in (\ref{subp2}), such that
\begin{equation}
\lim_{k\rightarrow 0}k^2\left( \rho^{(n)\alpha{\bar\alpha}}_{\gamma{\bar\gamma}1}(k)|_{\text{leading Casimir contribution}}
 -S_{QCD}(k)\rho^{(n)\alpha{\bar\alpha}}_{\gamma{\bar\gamma}0}\right) = 0,
\end{equation}
where we have now introduced the restriction to the leading
color Casimir terms at one-loop\footnote{These correspond with
maximally non-Abelian terms in Ref.~\cite{gatherall} but computed
exactly rather than in the eikonal approximation.} 
so that in the expression for the respective one-loop correction 
$\rho^{(n)}_1$ and in that for  
for $S_{QCD}(k)$ given in Refs.~\cite{delaney}, only the
terms proportional to $C_F$ should be retained here as
we focus on the f\=f$\rightarrow$f\=f case, where f denotes a fermion.
(Henceforth, when we refer to $k\rightarrow 0$ gluons we are always referring
for virtual gluons to the corresponding regime of the 4-dimensional
loop integration in the computation of $ M^{(n)}_\ell$.)

In Ref.~\cite{delaney},
the respective authors have calculated $S_{QCD}(k)$
using the running quark masses to regulate its collinear
mass singularities, for example;  n-dimensional regularization of
the 't Hooft-Veltman type is also possible for
these mass singularities and we will also illustrate this presently.

We stress that $S_{QCD}(k)$
has a freedom in it corresponding to the fact that any
function $\Delta S_{QCD}(k)$ which has the property that
$\lim_{k\rightarrow 0} k^2\Delta S_{QCD}(k)\rho^{(n)}_0=0$
may be added to it.

Since the virtual gluons in $\rho^{(n)}_{\ell}$
are all on equal footing by the symmetry of this function,
if we look at gluon $\ell$, for example, we may write
, for $k_{\ell} \rightarrow (0,0,0,0)\equiv O$ while the
remaining $k_i$ are fixed away from $O$, the representation
\begin{equation}
 \rho^{(n)}_\ell = S_{QCD}(k_\ell)*\rho^{(n)}_{\ell-1}(k_1,\cdots,k_{\ell-1})
 +\beta^1_\ell(k_1,\cdots,k_{\ell-1};k_\ell)
\label{subp3}
\end{equation}
where the residual amplitude $\beta^1_\ell(k_1,\cdots,k_{\ell-1};k_\ell)$
will now be taken as defined by this last equation. It has two nice 
properties:
\begin{itemize} 
\item it is symmetric in its first $\ell - 1$ 
arguments 
\item the
IR singularities for gluon $\ell$ {\it that are contained in 
$S_{QCD}(k_\ell)$} are no longer contained in it. 
\end{itemize}

We do not at this point
discuss the extent to which there are any further remaining IR singularities
for gluon $\ell$ in $\beta^1_\ell(k_1,\cdots,k_{\ell-1};k_\ell)$.
In an Abelian gauge theory like QED, as has been shown by Yennie, Frautschi
and Suura in Ref.~\cite{yfs},
there would not be any further
such singularities; for a non-Abelian gauge theory like QCD, this point
requires further discussion and 
we will come back to this point presently.

We rather now stress that
if we apply the representation (\ref{subp3}) again we may write
{\small
\begin{eqnarray}
   \rho^{(n)}_\ell = S_{QCD}(k_\ell)S_{QCD}(k_{\ell-1})* \rho^{(n)}_{\ell-2}(k_1,\cdots,
 k_{\ell-2})\nonumber \\
 + S_{QCD}(k_\ell)\beta^1_{\ell-1}(k_1,\cdots,k_{\ell-2};k_{\ell-1})
\nonumber \\
+ S_{QCD}(k_{\ell-1})\beta^1_{\ell-1}(k_1,\cdots,k_{\ell-2};k_\ell)\nonumber \\
+ \beta^2_\ell(k_1,\cdots,k_{\ell-2};k_{\ell-1},k_\ell),
\label{subp4}
\end{eqnarray} 
where this last equation serves to define the function
$\beta^2_\ell(k_1,\cdots,k_{\ell-2};k_{\ell-1},k_\ell)$. It has two nice
properties:
\begin{itemize} 
\item it is symmetric in its first $\ell -2$ arguments
and in its last two arguments $k_{\ell-1},k_\ell$ 
\item the infrared
singularities for gluons $\ell-1$ and $\ell$ that are contained in
$S_{QCD}(k_{\ell-1})$ and $S_{QCD}(k_\ell)$ are no longer contained in
it. 
\end{itemize}

Continuing in this way, with repeated application of (\ref{subp3}),
we get finally the rigorous, exact rearrangement of the 
contributions to $\rho^{(n)}_\ell$ as
\begin{eqnarray}
\rho^{(n)}_\ell = S_{QCD}(k_1)\cdots S_{QCD}(k_\ell)\beta^0_0+\sum_{i=1}^\ell\prod_{j\neq i}
S_{QCD}(k_j)\beta^1_1(k_i) +\cdots \nonumber\\
+\beta^\ell_\ell(k_1,\cdots,k_\ell),
\label{subp5}
\end{eqnarray}
where the virtual gluon residuals  $\beta^i_i(k'_1,\cdots,k'_i)$
have two nice properties:
\begin{itemize} 
\item they are symmetric functions of their
arguments 
\item they do not contain any of the IR singularities which
are contained in the product 
$S_{QCD}(k'_1)\cdots S_{QCD}(k'_i)$.
\end{itemize}

Henceforth, we denote  
$\beta^i_i$ as the function $\beta_i$ for reasons of pedagogy.
We can not stress too much that (\ref{subp5}) is an {\it exact}
rearrangement of the contributions of the Feynman diagrams which
contribute to $\rho^{(n)}_\ell$; it involves no approximations.
Here also we note that the
question of the absolute convergence of these Feynman diagrams
from the standpoint of constructive field theory remains open as usual. Yennie,
Frautschi and Suura~\cite{yfs} have already stressed that Feynman diagrammatic
perturbation theory is non-rigorous from this standpoint.
What we do claim is that the relationship between the YFS expansion and the
usual perturbative Feynman diagrammatic expansion is itself rigorous
even though neither of the two expansions themselves is rigorous.
\par

Introducing (\ref{subp5}) into (\ref{subp1}) yields
a representation similar to that of YFS, and we
will call it a ``YFS representation'', 
\begin{equation}
{\cal M}^{(n)} = e^{\alpha_sB_{QCD}}\sum_{j=0}^\infty{\sf m}^{(n)}_j,
\label{yfsrepv}
\end{equation}
where we have defined
\begin{equation}
\alpha_s(Q)B_{QCD} = \int {d^4k\over (k^2-\lambda^2+i\epsilon)}S_{QCD}(k)
\label{vbfn}
\end{equation}
and
\begin{equation}
 {\sf m}^{(n)}_j = {1\over {j!}}\int\prod_{i=1}^j{d^4k_i\over k_i^2-\lambda^2+i\epsilon}
       \beta_j(k_1,\cdots,k_j). 
\label{irfreev}
\end{equation}
We say that (\ref{yfsrepv}) is similar to the respective result
of Yennie, Frautschi and Suura in Ref.~\cite{yfs}
and is not identical to it because we have not
proved that the functions $\beta_i(k_1,...,k_i)$ are completely
free of virtual IR singularities. What have shown is that they do not
contain the IR singularities in the product 
$S_{QCD}(k_1)\cdots S_{QCD}(k_i)$ so that
${\sf m}^{(n)}_j$ does not contain the virtual IR divergences
generated by this product when it is integrated over the respective
4j-dimensional j-virtual gluon phase space. In an Abelian gauge theory,
there are no other possible virtual IR divergences; in the non-Abelian
gauge theory that we treat here, such additional IR divergences
are possible and are expected; but, the 
result (\ref{yfsrepv}) does have an improved
IR divergence structure over (\ref{subp1}) in that all of the
IR singularities associated with $S_{QCD}(k)$ are explicitly
removed from the sum over the virtual IR improved loop contributions
${\sf m}^{(n)}_j$ to all orders in $\alpha_s(Q)$.\par

Turning now to the analogous rearrangement of the real IR singularities in
the differential cross section associated with the ${\cal M}^{(n)}$,
we first note that we may write this cross section as follows
according to the standard methods
\begin{eqnarray}
  d\hat\sigma^n = {e^{2\alpha_sReB_{QCD}}\over {n !}}\int\prod_{m=1}^n
{d^3k_m\over (k_m^2+\lambda^2)^{1/2}}\delta(p_1+q_1-p_2-q_2-\sum_{i=1}^nk_i)
\nonumber\\       
\bar\rho^{(n)}(p_1,q_1,p_2,q_2,k_1,\cdots,k_n)
{d^3p_2d^3q_2\over p^0_2 q^0_2},
\label{diff1}
\end{eqnarray}
where we have defined
\begin{equation}
\bar\rho^{(n)}(p_1,q_1,p_2,q_2,k_1,\cdots,k_n)=
\sum_{color,spin} \|\sum_{j=0}^\infty{\sf m}^{(n)}_j\|^2
\label{diff2}
\end{equation}}
in the incoming Q\=Q' cms system
and we have absorbed the remaining kinematical factors for
the initial state flux, spin and color averages into the
normalization of the amplitudes ${\cal M}^{(n)}$ for reasons of
pedagogy so that the $\bar\rho^{(n)}$ are averaged over initial spins
and colors and summed over final spins and colors.
We now proceed in complete analogy with the discussion
of $\rho^{(n)}_\ell$ above. \par

Specifically, 
for the functions $\bar\rho^{(n)}(p_1,q_1,p_2,q_2,k_1,\cdots,k_n)
\equiv \bar\rho^{(n)}(k_1,\cdots,k_n)$ which are symmetric functions
of their arguments $k_1,\cdots,k_n$, we define first, for $n=1$, 
\begin{equation}
\lim_{|\vec{k}|\rightarrow 0}\vec{k}^2\left(\bar\rho^{(1)}(k)|_{\text{leading Casimir contribution}}
-\tilde S_{QCD}(k)\bar\rho^{(0)}\right) = 0, 
\label{realS}
\end{equation}
where the real infrared function $\tilde S_{QCD}(k)$ is rigorously
defined by this last equation and is explicitly computed in
Refs.~\cite{delaney}, wherein we retain here only
the terms proportional to $C_F$ 
from the result in Ref.~\cite{delaney}
; like its virtual counterpart $S_{QCD}(k)$
it has a freedom in it in that any function $\Delta\tilde S_{QCD}(k)$
with the property that 
$\lim_{|\vec{k}|\rightarrow 0}\vec{k}^2\Delta\tilde S_{QCD}(k)=0$ 
may be added to it without affecting the defining relation (\ref{realS}).
 
We can again repeat the 
analogous arguments of Ref.~\cite{yfs}, following the
corresponding steps in (\ref{subp3})-(\ref{irfreev}) 
above for $S_{QCD}$ to get 
the ``YFS-like'' result {\small 
\begin{equation}
\begin{split}
d\hat\sigma_{\rm exp}&= \sum_n d\hat\sigma^n \\
         &=e^{\rm SUM_{IR}(QCD)}\sum_{n=0}^\infty\int\prod_{j=1}^n{d^3
k_j\over k^0_j}\int{d^4y\over(2\pi)^4}e^{iy\cdot(p_1+q_1-p_2-q_2-\sum k_j)+
D_\rQCD}\\
&*\bar\beta_n(k_1,\ldots,k_n){d^3p_2\over p_2^{\,0}}{d^3q_2\over
q_2^{\,0}}
\end{split}
\label{subp10}
\end{equation}
with 
\[ {SUM}_{IR}(QCD)=2\alpha_s ReB_{QCD}+2\alpha_s\tilde B_{QCD}(\Kmax),\]
\[ 2\alpha_s\tilde B_{QCD}(\Kmax)=\int{d^3k\over k^0}\tilde S_\rQCD(k)
\theta(\Kmax-k),\]
 \begin{equation} D_\rQCD=\int{d^3k\over k}\tilde S_\rQCD(k)
\left[e^{-iy\cdot k}-\theta(\Kmax-k)\right],\label{subp11}\end{equation}
\[{1\over 2}\bar\beta_0=d\sigma^{\rm(1-loop)}-2\alpha_s{\rm Re}B_\rQCD d\sigma_B,\]
\begin{equation}{1\over 2}\bar\beta_1=d\sigma^{B1}-\tilde S_\rQCD(k)d\sigma_B,\quad\ldots\label{subp12}\end{equation}
where the $\bar\beta_n$ are the QCD hard gluon residuals defined above; they
are the non-Abelian analogs of the hard photon residuals defined by YFS.
Here, for illustration, we have recorded the
relationship between the $\bar\beta_n$, $n=0,1$ through ${\cal O}(\alpha_s)$
and the exact one-loop and single bremsstrahlung cross sections,
$d\sigma^{\rm(1-loop)}$, $d\sigma^{B1}$, respectively, where the latter
may be taken from Ref.~\cite{qqOalphas}
We stress two things about the right-hand side of
(\ref{subp10}) :
\begin{itemize}
\item It does not depend on the dummy parameter $K_{max}$ which has been
introduced for cancellation of the infrared divergences in 
$SUM_{IR}(QCD)$ to all orders in $\alpha_s(Q)$ where $Q$ is the hard
scale in the parton scattering process under study here.
\item Its analog can also be derived in our new CEEX~\cite{ceex:2001} 
format.
\end{itemize}

We now
return to the property of (\ref{subp10}) that distinguishes it
from the Abelian result derived by Yennie, Frautschi and Suura 
-- namely, the fact
that, owing to its non-Abelian gauge theory origins, it is in general
expected 
that there are infrared divergences in the $\bar\beta_n$ which were not
removed into the $S_{QCD},\tilde S_{QCD}$ when these infrared functions
were isolated in our derivation of (\ref{subp10}).}\par

More precisely, the left-hand side of (\ref{subp10}) is the fundamental
reduced parton cross section and it should be infrared finite or else
the entire QCD parton model has to be abandoned. 

There is an observation
in the literature~\cite{chris} that 
unless we use the approximation of massless incoming quarks,
the reduced parton cross section
on the left-hand side of (\ref{subp10}) diverges in the infrared
regime at ${\cal O}(\alpha_s^2(Q))$. 
We do not
go into this issue here but either
use the quark masses strictly as collinear limit regulators
so that they are set to zero in the numerators of all
Feynman diagrams in such a way that the limit
$\lim_{m_q^2/E_q^2\rightarrow 0}$, where $E_q$ is the quark
energy, is taken everywhere that
it is finite or, alternatively, we use
n-dimensional methods
to regulate such divergences while setting the quark masses
to zero as that is an excellent approximation for the
light quarks at FNAL and LHC energies -- we take this issue up elsewhere.\par

From the infrared finiteness of the left-hand side of (\ref{subp10})
and the infrared finiteness of $SUM_{IR}(QCD)$, it follows
that the quantity 
\[d\bar{\hat\sigma}_{\rm exp}\equiv 
e^{\rm -SUM_{IR}(QCD)}d\hat\sigma_{\rm exp}\]
must also be infrared finite to all orders in $\alpha_s$.

As we assume the QCD theory makes sense in some neighborhood of the
origin for $\alpha_s$, we conclude that each order in $\alpha_s$
must make an infrared finite contribution to $d\bar{\hat\sigma}_{\rm exp}$. 
At ${\cal O}(\alpha_s^0(Q))$ , the only contribution to 
$d\bar{\hat\sigma}_{\rm exp}$ is the respective Born cross section
given by $\bar\beta^{(0)}_0$ in (\ref{subp10}) and it
is obviously infrared finite, where we use henceforth the notation
$\bar\beta^{(\ell)}_n$ to denote the ${\cal O}(\alpha_s^{\ell}(Q))$
part of $\bar\beta_n$. Thus, we conclude that
the lowest hard gluon residual $\bar\beta^{(0)}_0$ is infrared finite.

Let us now define the left-over 
non-Abelian infrared divergence part
of each contribution $\bar\beta^{(\ell)}_n$ via
\[ \bar\beta^{(\ell)}_n= \tilde{\bar\beta}^{(\ell)}_n + D\bar\beta^{(\ell)}_n\]
where the new function $\tilde{\bar\beta}^{(\ell)}_n$ is now completely
free of any infrared divergences and the function $D\bar\beta^{(\ell)}_n$
contains all left-over infrared divergences in $\bar\beta^{(\ell)}_n$
which are of non-Abelian origin and is normalized to vanish in the
Abelian limit $f_{abc}\rightarrow 0$ where $f_{abc}$ are the group
structure constants.

Further, we define $D\bar\beta^{(\ell)}_n$
by {\it a} minimal subtraction of the respective IR divergences in it
so that it only contains the actual pole and transcendental
constants, $1/\epsilon -C_E$ for $\epsilon=2-d/2$, where $d$ is
the dimension of space-time, in dimensional regularization
or $\ln \lambda^2$ in the gluon mass regularization. Here, $C_E$ is
Euler's constant. 

For definiteness, we 
write this out explicitly as follows:
\[\int dPh\; D\bar\beta^{(\ell)}_n\equiv \sum_{i=1}^{n+\ell} d^{n,\ell}_i\ln^i(\lambda^2)\] 
where the coefficient functions $d^{n,\ell}_i$ are independent of $\lambda$ for
$\lambda\rightarrow 0$ and $dPh$ is the respective n-gluon Lorentz
invariant phase space.

At ${\cal O}(\alpha_s^n(Q))$, the IR finiteness
of the contribution to $d\bar{\hat\sigma}_{\rm exp}$ then requires
the contribution {\small
\begin{eqnarray}
d\bar{\hat\sigma}^{(n)}_{\rm exp} \equiv
\int\sum_{\ell=0}^n\frac{1}{\ell!}\prod_{j=1}^{\ell}\int_{k_j\ge K_{max}}{d^3
k_j\over k_j}\tilde S_{QCD}(k_j)\sum_{i=0}^{n-\ell}\frac{1}{i!}\prod_{j=\ell+1}^{\ell+i}
\nonumber\\
\int{d^3k_j\over k^0_j}
\bar\beta^{(n-\ell-i)}_i(k_{\ell+1},\ldots,k_{\ell+i}){d^3p_2\over p_2^{\,0}}{d^3q_2\over
q_2^{\,0}}\label{subp13}
\end{eqnarray}}
to be finite. 

From this it follows that {\small
\begin{eqnarray}
Dd\bar{\hat\sigma}^{(n)}_{\rm exp} \equiv
\int\sum_{\ell=0}^n\frac{1}{\ell!}\prod_{j=1}^{\ell}\int_{k_j\ge K_{max}}{d^3
k_j\over k_j}\tilde S_{QCD}(k_j)\sum_{i=0}^{n-\ell}\frac{1}{i!}\prod_{j=\ell+1}^{\ell+i}
\nonumber\\
\int{d^3k_j\over k^0_j} D\bar{\beta}^{(n-\ell-i)}_i(k_{\ell+1},\ldots,k_{\ell+i}){d^3p_2\over p_2^{\,0}}{d^3q_2\over
q_2^{\,0}}\label{subp14}
\end{eqnarray}}
is finite. Since the integration region for the final particles is
arbitrary, the independent powers of the IR regulator $\ln(\lambda^2)$ in this
last equation must give vanishing contributions.
This means that
we can drop the $D\bar\beta^{(\ell)}_n$ from our result
(\ref{subp10}) because they do not make a net contribution to the final
parton cross section $\hat\sigma_{\rm exp}$. We thus finally arrive
at the new rigorous result{\small
\begin{equation}
\begin{split}
d\hat\sigma_{\rm exp}&= \sum_n d\hat\sigma^n \\
         &=e^{\rm SUM_{IR}(QCD)}\sum_{n=0}^\infty\int\prod_{j=1}^n{d^3
k_j\over k_j}\int{d^4y\over(2\pi)^4}e^{iy\cdot(p_1+q_1-p_2-q_2-\sum k_j)+
D_\rQCD}\\
&*\tilde{\bar\beta}_n(k_1,\ldots,k_n){d^3p_2\over p_2^{\,0}}{d^3q_2\over
q_2^{\,0}}
\end{split}
\label{subp15}
\end{equation}
where now the hard gluon residuals 
$\tilde{\bar\beta}_n(k_1,\ldots,k_n)$
defined by 
\begin{equation}
\tilde{\bar\beta}_n(k_1,\ldots,k_n= \sum_{\ell=0}^\infty 
\tilde{\bar\beta}^{(\ell)}_n(k_1,\ldots,k_n)
\label{newbeta}
\end{equation}
are free of all infrared divergences to all 
orders in $\alpha_s(Q)$.
This is a basic result of this Appendix.\par
 
We note here that, contrary to what was claimed in the Appendix of the
first paper in Refs.~\cite{delaney} and consistent with what is
explained in the third reference in ~\cite{delaney}, the arguments in 
the first paper in Refs.~\cite{delaney}
are not sufficient to derive the respective analog of eq.(\ref{subp15});
for, they did not really expose the compensation between
the left over genuine non-Abelian IR virtual and real singularities
between $\int dPh\bar\beta_n$ and $\int dPh\bar\beta_{n+1}$ respectively
that really distinguishes
QCD from QED, where no such compensation occurs in the $\bar\beta_n$
residuals for QED.\par

We point-out that the general non-Abelian exponentiation of the 
eikonal cross sections in QCD has been proven formally in 
Ref.~\cite{gatherall}. The contact between Ref.~\cite{gatherall}
and our result (\ref{subp15}) is that, in the language of
Ref.~\cite{gatherall}, our exponential factor corresponds to the 
N=1 term in the exponent of eq.(10) of the latter reference.
One also sees immediately the fundamental difference between what we
derive in (\ref{subp15}) and the eikonal formula in Ref.~\cite{gatherall}:
our result (\ref{subp15}) is an {\it exact} re-arrangement 
of the complete cross section
whereas the result in eq.(10) of Ref.~\cite{gatherall} is an
{\it approximation} to the complete cross section in which all terms that
could not be eikonalized and exponentiated have been dropped.
\par


\newpage
\begin{thebibliography}{99}
\bibitem{sm1}S.L. Glashow, Nucl. Phys. {\bf 22} (1961) 579; 
S. Weinberg, Phys. Rev. Lett. {\bf 19} (1967) 1264;
A. Salam, in {\em Elementary Particle Theory}, ed. N. Svartholm
(Almqvist and Wiksells, Stockholm, 1968), p. 367;
G.~'t Hooft and M.~Veltman, Nucl. Phys. {\bf B44},189 (1972)
and {\bf B50}, 318 (1972); 
G.~'t Hooft, {\it ibid.} {\bf B35}, 167 (1971); M.~Veltman, {\it ibid.} {\bf B7}, 637 (1968); 
\bibitem{qcd}
D. J. Gross and F. Wilczek, 
Phys. Rev. Lett. {\bf 30} (1973) 1343;
H. David Politzer, {\it ibid.}{\bf 30} (1973) 1346; see also
, for example, F. Wilczek, in {\em Proc. 16th International Symposium on Lepton and 
Photon Interactions, Ithaca, 1993}, eds. P. Drell and D.L. Rubin 
(AIP, NY, 1994) p. 593, and references therein.
\bibitem{dglap}
G. Altarelli and G. Parisi, Nucl. Phys. {\bf B126} (1977) 
298; Yu. L. Dokshitzer, Sov. Phys. JETP {\bf 46} (1977) 641;
L.~N. Lipatov, Yad. Fiz. {\bf 20} (1974) 181; V. Gribov and L. Lipatov,
Sov. J. Nucl. Phys. {\bf 15} (1972) 675, 938; see also J.C. Collins and J. Qiu,
Phys. Rev. D{\bf 39} (1989) 1398 for an alternative discussion
of the lowest order DGLAP theory.
\bibitem{cteq} S. Kretzer, H.L. Lai, F.I. Olness, W.K. Tung, MSU-HEP-030101, BNL-NT-03-2, RBRC-325; Phys. Rev. D{\bf 69} (2004) 114005, and references therein.
\bibitem{mrst} A.D. Martin, R.G. Roberts, W.J. Stirling and R.S. Thorne,
preprint IPPP-04-62, DCPT-04-124, CAVENDISH-HEP-2004-28;
Eur. Phys. J. C{\bf 39} (2005) 155, and references therein.
\bibitem{greya} M. Gluck, E. Reya, M. Stratmann and W. Vogelsang, preprint DO-TH-2000-14, RBRC-148, TPR-00-21, Phys. Rev. D{\bf 63} (2001) 094005.
\bibitem{other-pdfs} See, for example, Robert S. Thorne, A.D. Martin, R.G. Roberts and W.J. Stirling,  CAVENDISH-HEP-05-12, hep-ph/0507015, in {\it Proceedings of 13th International Workshop on Deep Inelastic Scattering (DIS 05), Madison, Wisconsin, 27 Apr - 1 May 2005}, to appear.
\bibitem{field} R.D. Field, {\it Applications of Perturbative QCD},(Addison-Wesley Publ. Co., Inc, Redwood City, 1989).
\bibitem{yellowbook}F. Berends et al., ''Z Line Shape'', in {\it Z Physics at LEP 1, v. 1}, CERN-89-08, eds. G. Altarelli, R. Kleiss, and C. Verzegnassi,(CERN, Geneva, 1989) p. 89, and references therein.
\bibitem{jsw} S. Jadach, M. Skrzypek and B.F.L. Ward, Phys. Rev. D{\bf 47} (1993) 3733; Phys. Lett. B{\bf 257} (1991) 173; in ''$Z^o$ Physics'', {\it Proc.
XXVth Rencontre de Moriond, Les Arcs, France, 1990}, ed. J. Tran Thanh Van 
(Editions Frontieres, Gif-Sur-Yvette, 1990); S. Jadach et al., Phys. Rev. D{\bf 44} (1991) 2669.
\bibitem{jw} S. Jadach and B.F.L. Ward, preprint TPJU 19/89;
in {\it Proc. Brighton Workshop}, eds. N. Dombey and F. Boudjema (Plenum, London, 1990), p. 325.
(1991) 577.
\bibitem{berge} C. Balazs et al., Phys. Lett. {\bf B637} (2006) 235.
\bibitem{cdf1} D. Acosta et al., (CDF Collaboration), Phys. Rev. Lett. {\bf 95} (2005) 022003.
\bibitem{alball} S. Albino and R. Ball, {\sl Phys. Lett. B}{\bf 513}(2001) 93.
\bibitem{mikhlv} S.V. Mikhailov, {\it Phys. Lett. B}{\bf 431} (1998) 387.
\bibitem{qcdexp} B.F.L. Ward and S. Jadach, {\it Acta Phys.Polon.} 
{\bf B33} (2002)
1543; in {\it Proc. ICHEP2002},
ed. S. Bentvelsen {\it et al.},( North Holland, Amsterdam, 2003 ) p. 275
; B.F.L. Ward and S. Jadach, {\sl Mod. Phys. Lett.}{\bf A14} (1999) 491
; D. DeLaney {\it et al.}, {\sl Mod. Phys. Lett.} {\bf A12} (1997) 2425; 
C. Glosser, S. Jadach, B.F.L. Ward and S.A.,{\sl Mod. Phys. Lett.}{\bf A 19}(2004) 2113; B.F.L. Ward, C. Glosser, S. Jadach and S.A. Yost, in {\it Proc. DPF 2004}, Int. J. Mod. Phys. {\bf A20} (2005) 3735; in {\it Proc. ICHEP04, vol. 1}, eds. H. Chen et al.,(World. Sci. Publ. Co., Singapore, 2005) p. 588; B.F.L. Ward and S. Yost, preprint BU-HEPP-05-05, and references therein.
\bibitem{delaney}
D. DeLaney {\it et al.},Phys. Rev. D{\bf 52} (1995) 108; Phys. Lett. {\bf B342} (1995) 239; Phys. Rev. D{\bf 66} (2002) 019903(E).
\bibitem{yfs}D.~R.~Yennie, S.~C.~Frautschi, and H.~Suura, Ann. Phys. {\bf 13} (1961) 379;\newline
see also K.~T.~Mahanthappa, {\sl Phys.~Rev.~\bf 126} (1962) 329, for a related analysis.
\bibitem{qcdfactorzn} R.K. Ellis {\em et al.}, {\sl Phys. Lett.}{\bf B78} (1978) 281; {\sl Nucl.Phys.} {\bf B152} (1979) 285; D. Amati, R. Petronzio and G. Veneziano, {\it ibid.}{\bf B146} (1978) 29; S. Libby and G. Sterman, {\sl Phys. Rev.} {\bf D18} (1978) 3252; A. Mueller, {\it ibid.} {\bf D18} (1978) 3705.
\bibitem{fyodor} S. G. Gorishnii, S.A. Larin, and F. V. Tkachov, {\sl Phys. Lett.}{\bf B124} (1983) 217; 
S. G. Gorishnii and S.A. Larin, {\sl Nucl. Phys. B}{\bf 283} (1987) 452. 
\bibitem{elswh} B.F.L. Ward, to appear.
\bibitem{qcd1} H. Georgi and H. D. Politzer, Phys. Rev. D{\bf 9} (1974) 416; D. Gross and F. Wilczek, {\it ibid.}{\bf 9} (1974) 980.
\bibitem{siggi} S. Bethke, Nucl. Phys. Proc. Suppl.{\bf 135} (2004) 345.
\bibitem{carli} T. Carli {\em et al.}, in {\it Proc. HERA-LHC Workshop}, CERN-2005-014, eds. A. De Roeck and H. Jung,(CERN, Geneva, 2005) p. 78. 
\bibitem{mvermn} E.G. Floratos, D.A. Ross, C. T. Sachrajda, {\it Nucl.Phys.} {\bf B129}(1977) 66;{\it ibid.}{\bf B139}(1978) 545;
{\it ibid.}{\bf B152} (1979) 493,1979; A. Gonzalez-Arroyo, C. Lopez and F.J. Yndurain, {\it Nucl. Phys.}{\bf B153} (1979) 161; A. Gonzalez-Arroyo and C. Lopez,
{\it Nucl. Phys.} {\bf B166} (1980) 429; G. Curci, W. Furmanski and R. Petronzio, {\it Nucl. Phys.} {\bf B175} (1980) 27;  W. Furmanski and R. Petronzio, {\it Phys. Lett.} {\bf B97} (1980) 437;
E.G. Floratos, C. Kounnas and R. Lacaze, {\it Nucl. Phys.}B192:417,1981 ;
R. Hamberg and W. Van Neerven, {\it Nucl. Phys.} {\bf B379} (1992) 143.
\bibitem{mvovermn} S. Moch, J.A.M. Vermaseren and A. Vogt, Nucl. Phys. B688 (2004) 101; {\it ibid.} {\bf B691} (2004) 129, and references therein.
\bibitem{moment} G. Corcella and L. Magnea, preprint CERN-PH-TH-2005-090, DFTT-15-2005, Jun 2005; hep-ph/0506278, and references therein. 
\bibitem{sterman}G. Sterman,{\it Nucl. Phys.}B {\bf 281}, 310 (1987).
\bibitem{cattrent} S. Catani and L. Trentadue,
{\it Nucl. Phys.}B {\bf 327}, 323 (1989); {\it ibid.} {\bf 353}, 183 (1991). 
\bibitem{gps} S. Jadach, B.F.L. Ward and Z. Was, {\it Eur. Phys. J. }{\bf C22} (2001) 423.
\bibitem{ceex:2001}
S. Jadach, B.F.L. Ward and Z. Was, Phys. Rev. D{\bf 63} (2001) 113009.
\bibitem{tHvelt}G.~'t Hooft and M.~Veltman, Nucl. Phys. {\bf B44} (1972) 189 
and {\bf B50} (1972) 318, and references therein.
\bibitem{gatherall} J.G.M. Gatherall, Phys. Lett. B{\bf 133} (1983) 90.
\bibitem{qqOalphas} R. K. Ellis and J.C. Sexton, 
Nucl. Phys. B{\bf 269} (1986) 445.
\bibitem{chris}C. Di'Lieto, S. Gendron, I.G. Halliday and Christopher T. Sachrajda, Nucl. Phys. {\bf B183} (1981) 223;  S. Catani {\it et al.}, {\em ibid.} {\bf B264} (1986) 588 ; S. Catani, 
Z. Phys. {\bf C37} (1988) 357, and references therein.
\end{thebibliography}
\end{document}